\documentclass[11pt]{article}
\usepackage{hyperref}
\usepackage[blocks]{authblk}

\pdfoutput=1
\begin{document}
\title{Visualization of flow over a golf ball at Re = 110,000.}

\author[1]{Clinton Smith}
\author[2]{Nikolaos Beratlis}
\author[2]{Elias Balaras}
\author[1]{Kyle Squires}
\author[3]{Masaya Tsunoda}

\affil[1]{Arizona State University}
\affil[2]{University of Maryland} 
\affil[3]{SRI RD Ltd.}
        
\maketitle
\begin{abstract}

The drag on a golf ball can be reduced by dimpling the surface. There
have been few studies, primarily experimental, that provide
quantitative information on the details of the drag reduction
mechanisms.  To illuminate the underlying mechanisms, Direct Numerical
Simulation (DNS) is applied to the flow around a golf ball using an
immersed boundary method. Computations are performed using up to 500
processors on a range of mesh resolutions from 61 million points to
1.2 billion points.  Results are presented from simulations performed
at a Reynolds number of Re = $1.1\times 10^5$ using a grid of 1.2
billion points.  This video shows the development of instabilities in
the near-surface flow, as well as the delay of complete separation due
to the development of local shear layers that lead to local separation
and reattachment within individual dimples.

\end{abstract}

\section{Introduction}

This fluid dynamics video shows the evolution of the flow over a golf
ball from different several perspectives in space.  The variables
shown in the video are isosurfaces of the Q-criterion [1] colored by
spanwise vorticity, contours of spanwise vorticity, and an isosurface
of the vorticity magnitude.

The first section of the video presents an isosurface of the
Q-criterion colored by the spanwise vorticity, where white and blue
represent positive and negative values of the spanwise vorticity.  As
the view moves closer to the golf ball, several ``trains'' of vortices
become apparent near the surface about 45 degrees from the stagnation
point on the front of the ball.  These vortices are related to the
geometry of the surface as the generation of the structures follows
the lines of the dimples in the streamwise direction.

The Q-criterion isosurface is faded out to reveal planar contours of
the spanwise (out-of-plane) vorticity.  As the flow evolves from the
front to the rear of the dimples, local detachment leads to flow
structures generated by the developing shear layer.  The different
contour planes shown suggest the local detachment region within
individual dimples varies with the streamwise direction.

The contours of the spanwise vorticity are faded out and replaced by a
transparent isosurface of the vorticity magnitude.  Here, the vortical
structures in the flow are demonstrated to have streamwise and
spanwise dependence, and the effect of the dimples is apparent in the
generation of vortices in individual dimples.  As the observer moves
away from the golf ball, the effect of the local vortex generation by
the dimples is evident in the near-wake region.

The video animation file can be found at
\href{http://hdl.handle.net/1813/11586}{golfball\_movie\_cropped720x480\_qmax500k.mpg}

\section{References}
Hunt, J., Wray, A., Moin, P., ``Eddies, stream, and convergence zones in turbulent
flows'', {\em Center for Turbulence Research Report}, CTRS88, 1988.

\end{document}